\documentclass[11pt,preprint2]{aastex}
\usepackage{natbib}

\begin{document}

\title{An Observation of the Intermediate Polar XY Arietis with Chandra}

\author{A. Salinas\altaffilmark{1,2,3} \& E. M. Schlegel\altaffilmark{2}}

\altaffiltext{1}{University of Texas, Austin, TX}
\altaffiltext{2}{Harvard-Smithsonian Center for Astrophysics, Cambridge, MA}
\altaffiltext{3}{NSF REU student}

\begin{abstract}
Chandra serendipitously observed the eclipsing (80$^{\circ} <$ i $<
84^{\circ}$) intermediate polar, XY Arietis (=H0253+193), in two
separate but continuous observations five weeks apart.  XY Ari was in
a quiescent state during both observations.  We pursue the study of
phase-resolved spectra for this system focusing on the Fe K lines.
From the combined and separate data sets, we readily detect emission
lines of iron near 6.4, 6.7, and 6.9 keV at better than 99\%
significance in contrast to previous results.  We confirm the
orbit-phased sinusoidal absorption column behavior first observed with
Ginga as well as a sinusoid-like behavior as a function of spin phase.
The presence of the 6.4, 6.7, and 6.9 keV lines requires different
ionization states with $\xi$ $<$2 (6.4 keV) and ${\xi}$ $\sim$3.5-4
(6.7 and 6.9 keV) that must vary with phase.  We also detect emission
lines at 3.25, 4.8, and 5.4 keV that are not instrumental in origin.
The 4.8 keV line may be identified as Ca XIX (4.832 keV) and the 3.25
keV line may be Ar I K, but the 5.4 keV line has no obvious
identification.
\end{abstract}


\section{Introduction}

 XY Ari belongs to the class of binary stars known as cataclysmic
variables (CV's) which are characterized by mass transfer between a
red dwarf and a white dwarf \citep{warn95}.  The infalling mass may
create an accretion disk surrounding the white dwarf depending on the
strength of the magnetic field of the primary star. CV's in which the
primary has little to no magnetic field have large accretion disks;
systems for which the white dwarf has a large magnetic field (polars
or AM Her stars) do not have an accretion disk and matter accretes
onto one pole of the primary. XY Ari is a member of the intermediate
polar subclass (=IP or DQ Her stars; \citealt{patterson1994}).  These
stars may have both a magnetic field and an accretion disk, but debate
exists on the degree of disk versus diskless accretion (e.g.,
\citealt{norton1997}).  In either scenario, the mass transfer material
accretes onto the magnetic poles of the white dwarf; the differences
in the scenarios lie in the type of the accretion, whether arc-like
(disk-fed) or funnel-like (stream-fed) and the consequently different
observational signatures expected from the two methods (e.g.,
\citealt{norton1996} for power spectra).

XY Ari is the only fully eclipsing intermediate polar
known\footnote{Although Singh \& Rana (2003, A\&A, 410, 231) argue
that the eclipsing CV V1432 Aql is an IP; if confirmed, our use of
``only'' would be no longer correct.}; as luck would have it, the CV
is hidden behind a molecular cloud, Lynds 1457, and is virtually
invisible in the optical (visual extinction of $A_v=11.5$ $\pm$ 0.3;
\citealt{LDM}).  \cite{Kamata1991} determined XY Ari's X-ray spin
period, $\omega$, = 206.298 $\pm0.001 s$, and its orbital period
$\Omega$, to be 21,829 $\pm$3 s ($\approx$ 6.02 h), thereby firmly
establishing its classification as an IP. \cite{Hellier1997} observed
XY Ari during its only known outburst and placed a limit on the
accretion region of $<$0.002 of the white dwarf's surface area, as
well as constraining the inclination of the system to be
$80^\circ<i<84^\circ$.

The continuum emission of these systems provides the overall
properties (e.g., inclination, eclipse width, column absorption,
etc.), while the line emission has the potential to provide specific
details (e.g., temperatures, electron densities, emitting region
visibility). Of particular interest are the emission lines of iron in
the 6-7 keV band.  Previous studies of the lines studied their
behavior over the entire observation at CCD resolution
(\citealt{EzukaIshida1998}, \citealt{Osborne}) or at grating
resolution (e.g., \citealt{mukai2003}).  Previous time-resolved
studies of the X-ray emission focused solely on the continuum (eg.,
\citealt{Kamata93,norton1989}).  In our study of XY Ari using archival
Chandra data, we focus on the time-resolved behavior of the iron lines
in the 6.4-6.9 keV region.

\section{Observation} \label{obssect}

Chandra observed XY Ari on 2000 July 9-10 for approximately 56 ksec
and then again on 2000 August 17 for approximately 44 ksec (obs id:
943, target name: MBM12).  Chandra observations of length less than
$\sim$150 ksec are usually scheduled in one continuous pass at the
object; the July 2000 observation was interrupted by high solar
radiation and the remainding time was re-scheduled for mid-August.  A
few ksec were trimmed off the end of the first observation by the
automated data processing because of violations of background count
rates. The data were completely re-processed during an overall CXC
re-processing effort in 2001-2002 that improved various
calibration-related issues.  We obtained the re-processed data from
the archive.

Both observations placed XY Ari on the Advanced CCD Imaging
Spectrometer (ACIS)-S3 chip; the aim point on the S3 chip
serendipitously placed XY Ari $\sim$5.6 arcmin off-axis.  The off-axis
position mitigates the effects of event pile-up not only from the
reduced effective area ($\sim$86\%, \citealt{PropGuide}) but also from
the broader point spread function (=PSF; $\sim$2 times larger than
on-axis as defined by the 90\% encircled energy behavior).  To verify
that the source was free of pile-up, we generated a radial profile of
the source for the entire $\sim$100 ks and compared it to the
theoretical PSF.  For the profiles, we centered a series of 30 annuli
0.49 arcseconds apart on the source and sampled the background with an
annulus of inner radius 16 arcseconds and outer radius of 30
arcseconds.  We used the CIAO (Chandra Imaging and Analysis of
Observations v2.2.1)\footnote{http://asc.harvard.edu/ciao/} tools to
extract the counts in each annulus to construct the profile.  For the
theoretical PSF, we used the {\tt mkPSF} tool in CIAO and then
normalized the result by the total number of counts in the source
region. Both the theoretical and observed profiles agreed to within
0.4\%.  Agreement between the profiles, particularly in and near the
center, indicates little or no pileup exists in the source.

The observations of XY Ari, although separated in time, occurred
within a few weeks of each other.  As a result, the time-dependent
gain changes described in \cite{vikhlinin}\footnote{The web address
for this reference is
http://hea-www.harvard.edu/{$\sim$}alexey/acis/tgain/.} do not alter
the energy scales assigned to the two observations as the differences
in the gains between the two observations were insignificant.

\begin{figure*}
\begin{center}
\caption{\small Lightcurves for the (top) first and (bottom) second
observations of XY Ari with Chandra in the 0.3-10 keV band. Each bin
is 90s in size. The first observation included a 1.2 ksec time span
that was contaminated by a background-induced flare near 7959e7 that
was excised from the data.  The other four, broader drops in count
rate are the XY Ari eclipses.  Time is given in Chandra time, which is
seconds from MJD=50814.}
\label{fig1}
\scalebox{0.5}{\rotatebox{0}{\includegraphics{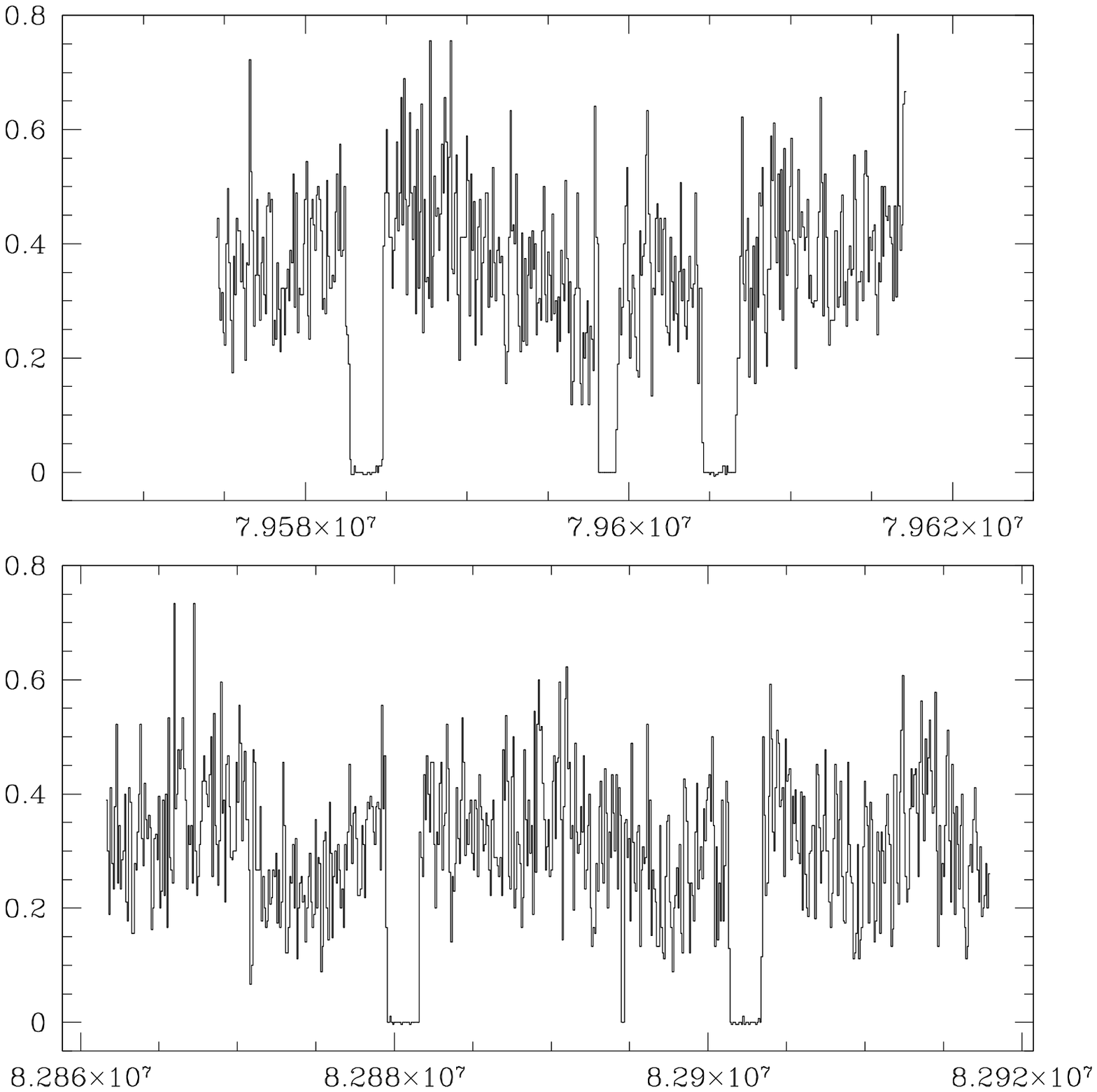}}}
\end{center}
\end{figure*}

Shown in Figure~\ref{fig1} are light curves for both observations
using bins of width 90 s in the energy range of 0.3-10 keV.  We
removed 1.2 ksec of time from the first observation in which the
background significantly increased, most likely caused by an increase
in the charged particle background induced by a solar flare.  The
times were corrected to the solar barycenter using the CIAO code {\tt
axbary} with the DE405 solar system ephemeris (consistent with the
ICRS reference system).  We then determined the source count rates for
both observations by centering a circle of radius of 0.14 arcminutes
at the source position.  Background counts were accumulated from an
annulus surrounding the source using an inner radius of 0.27
arcminutes and an outer radius of 0.54 arcminutes.  The source had a
count rate of 0.346 counts s$^{-1}$ (flare-subtracted data) with a
background rate of 0.0093 counts s$^{-1}$ during the first observation
and source and background rates of 0.302 counts s$^{-1}$ and 0.0096
counts s$^{-1}$, respectively, during the second observation.

\subsection{Orbit Period}

Two full orbital cycles of XY Ari are covered by the two observations.
We folded the light curve on the orbital period using the ephemeris of
\cite{Allan96}.  Within the errors, the predicted values match the
observed eclipse times.  We do not show the orbital light curve
because most of the features may be inferred from Figure~\ref{fig1}.
Table~\ref{ecltime} lists the mid-eclipse times determined from the
half-point at 50\% of the eclipse depth.  We also examined the
hardness ratios phased at the orbital period; we do not present the
ratios here since they are constant within the errors over the orbital
period.

\subsection{Spin Period}

\begin{figure*}
\begin{center}
\caption{\small Spin period-folded lightcurves folded on the 206.298s
spin period for both observations. Also shown are the 4-8 kev/1-4 kev
hardness ratios. The data have been repeated twice for clarity.}
\label{fig2}
\scalebox{0.4}{\rotatebox{0}{\includegraphics{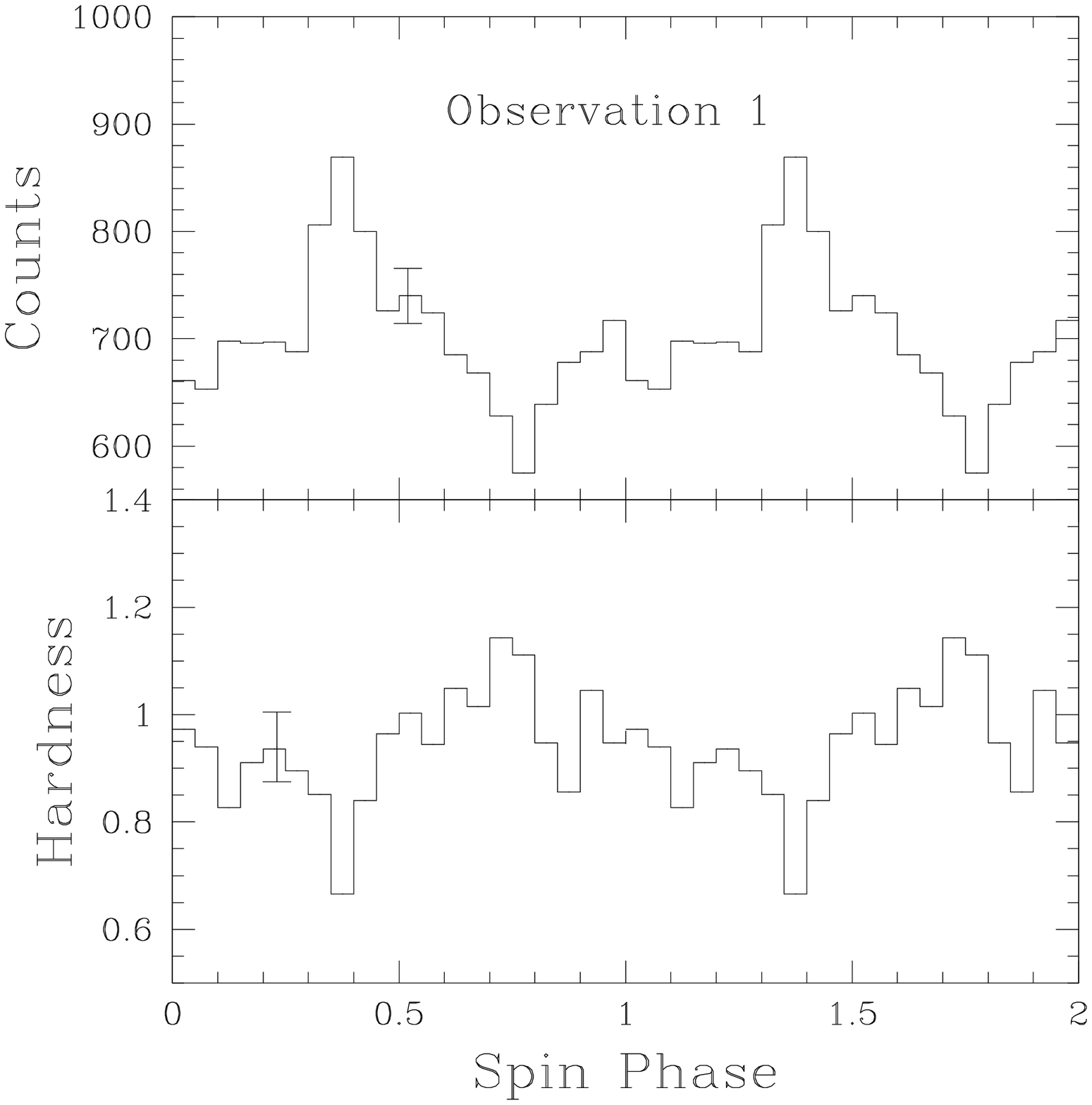}}}
\scalebox{0.4}{\rotatebox{0}{\includegraphics{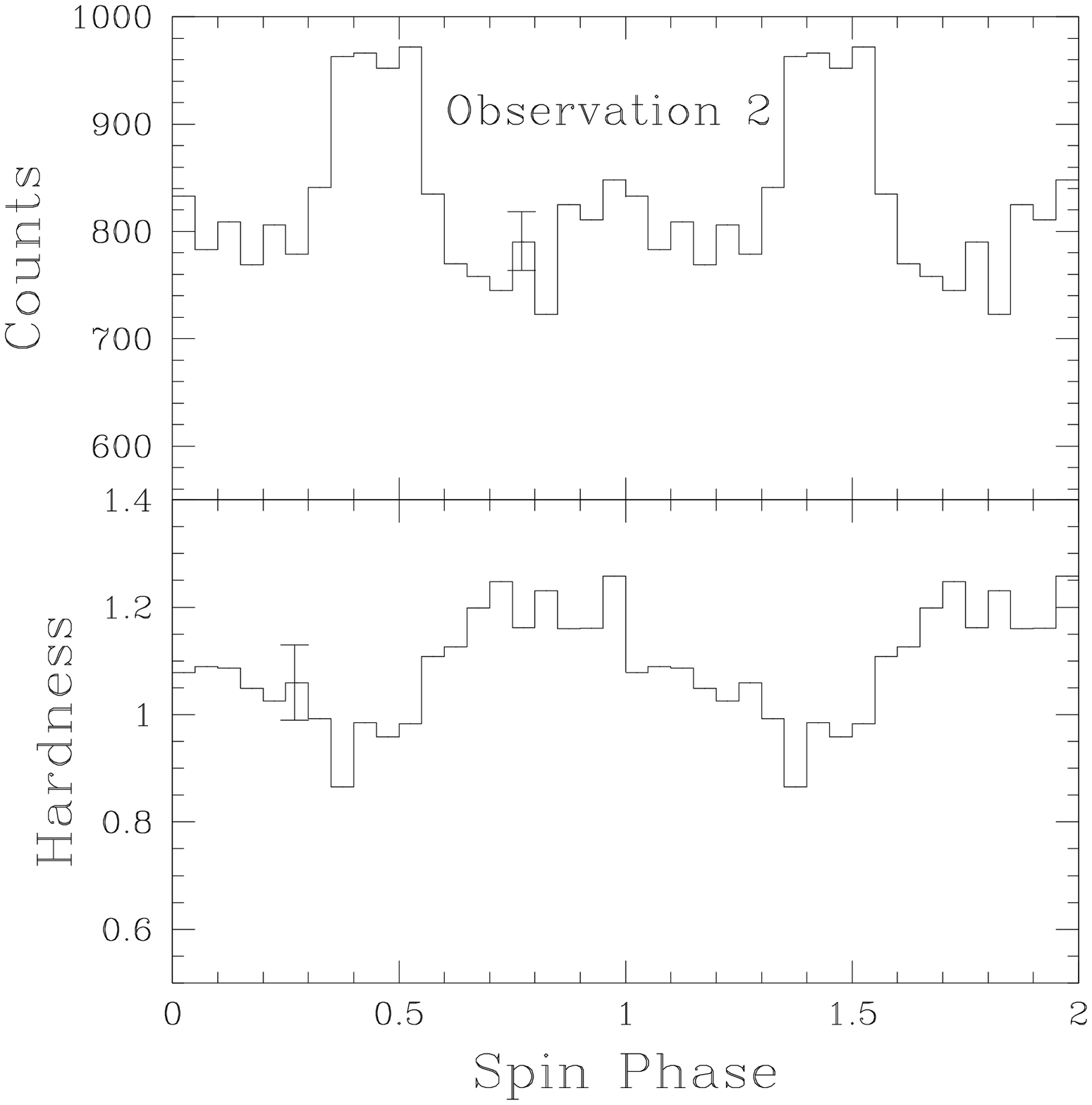}}}
\end{center}
\end{figure*}

Hellier (1997) used an ephemeris that started at TDB 245 0277.40382
and adopted the Kamata et al. spin period.  We adopted the same
ephemeris and folded the data on that ephemeris as shown in
Figure~\ref{fig2}.  A hardness ratio, defined as HR =
4-8~keV~/~1-4~keV was generated for each observation and is displayed
in the lower portion of Figure~\ref{fig2}.  A representative error bar
is also included.  The two observations are similar, but the width of
the pulse peak is significantly broader in the second observation
(0.15 vs. 0.20).

Previous observations of XY Ari by \cite{Hellier1997} studied the spin
pulse profile, and the spin pulse profiles in this observation
resemble those results.  The absolute spin pulses in the Chandra
observation are offset from earlier lightcurves by $\approx$0.32 phase
shift in the first observation, and by $\approx$0.30 phase shift in
the second observation.  These offsets most likely originate with an
error in the extrapolated ephemeris\footnote{See \cite{Hellier1997}
for more discussion on the error in the ephemeris.}.  The profiles are
essentially unchanged in shape which, for our purposes, is more
important.  We attempted to improve the emphemeris, but could not be
certain of the cycle count.

\subsection{Spin-Orbit behavior}

\begin{figure*}
\begin{center}
\caption{Orbit-phase data filtered into 3 phase bins for which each
bin is folded on the spin period.  Both the soft (0.5-4 keV) and hard
(4-8 keV) bands are shown. The data are repeated twice for clarity.}
\label{fig4}
\scalebox{0.4}{\rotatebox{0}{\includegraphics{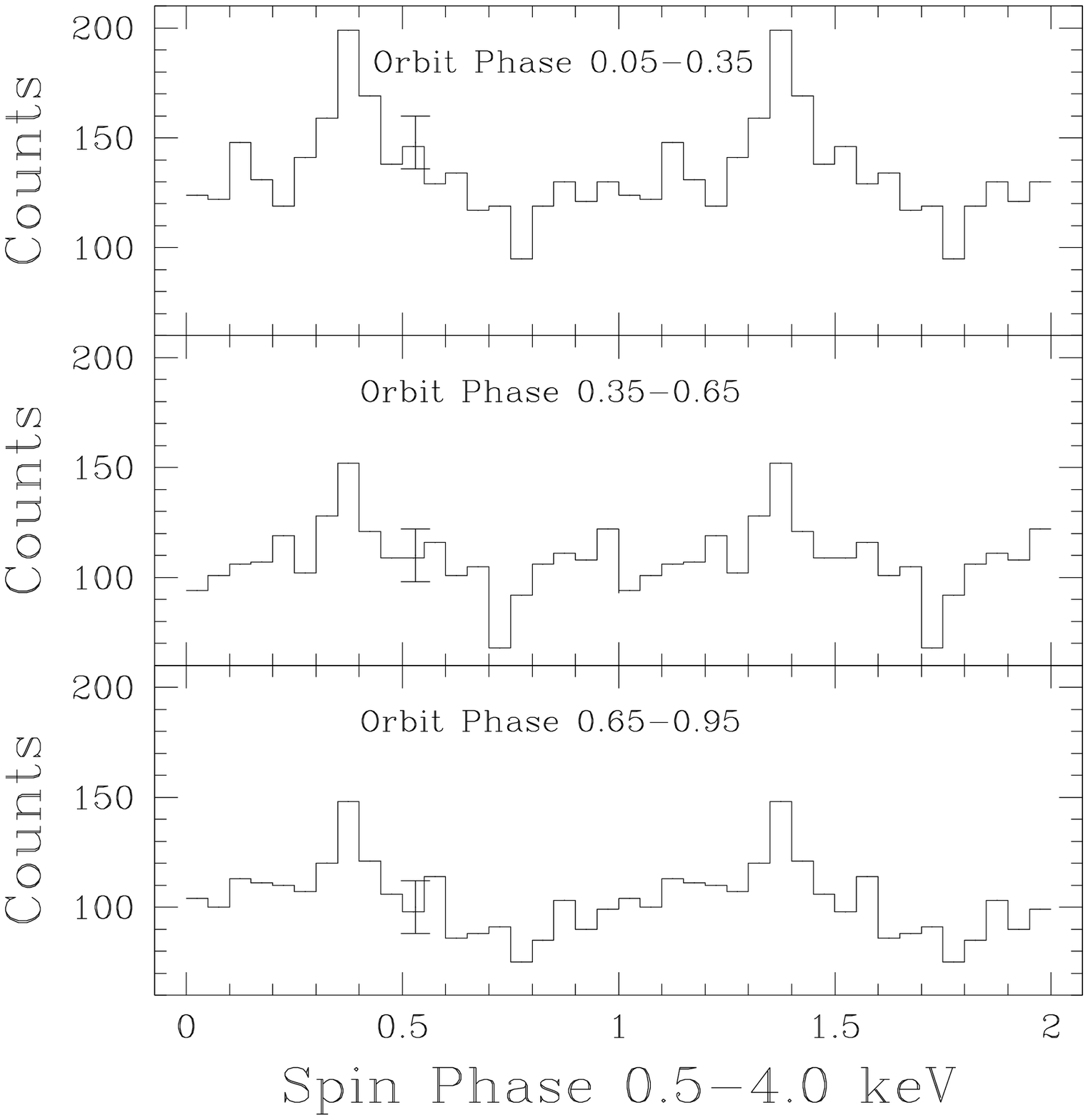}}}
\scalebox{0.4}{\rotatebox{0}{\includegraphics{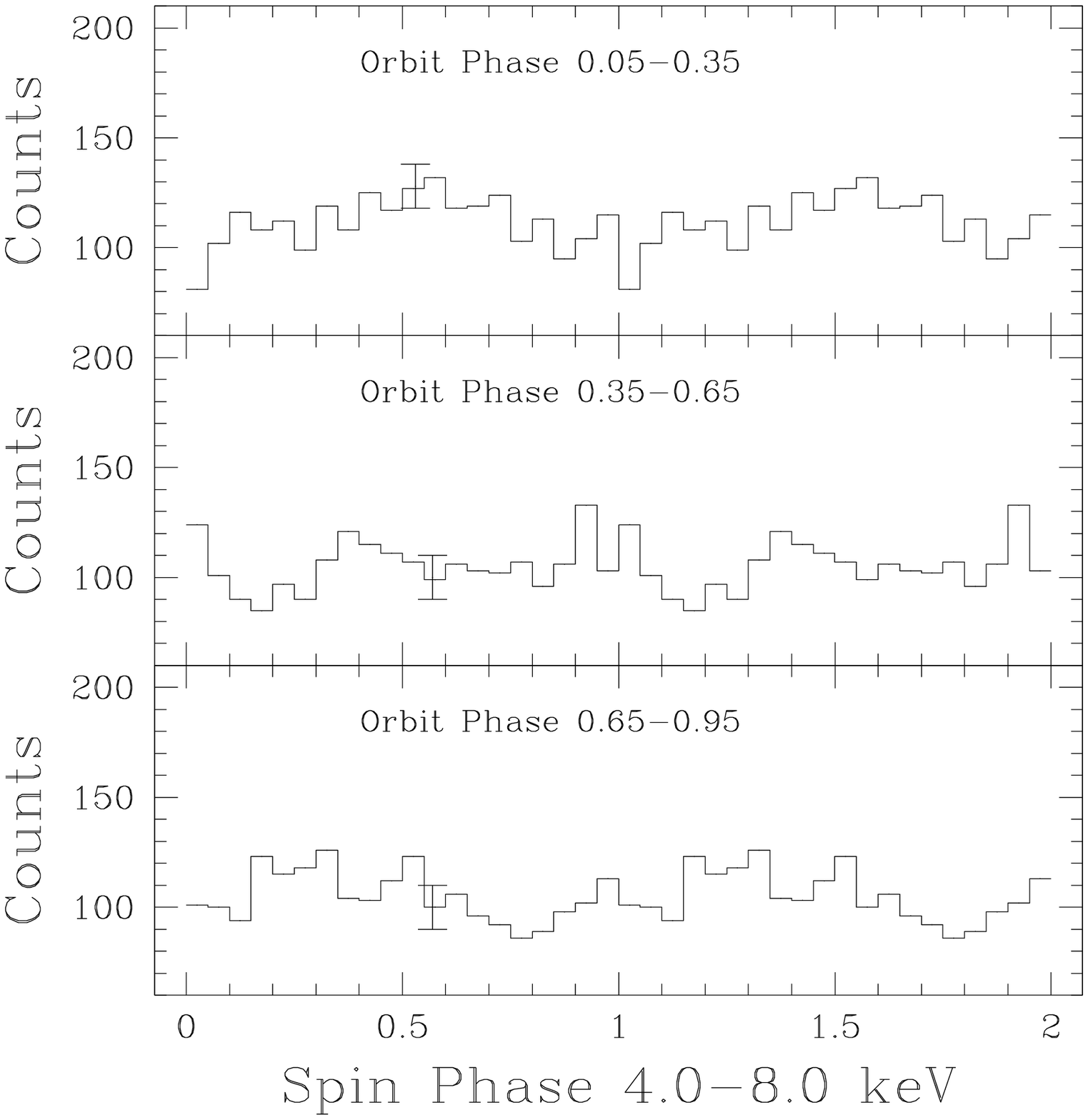}}}
\end{center}
\end{figure*}

Shown in Figure~\ref{fig4} are light curves, filtered into 3
orbit-phased bins, folded on the spin period, and separated into 2
energy bands (0.5-4.0 and 4.0-8.0 keV).  Phases immediately at or
adjacent to the eclipse were not included.  We note the peak near spin
phase 0.4 that is visible in the soft band, but not the hard band.
The hard band light curves are essentially consistent with constant
emission at all spin and orbit phases.

The count rate in the soft band (0.5-4.0 keV) during orbital phases
0.05-0.35 lies $\sim$10-15\% above the count rates in the other bins.
The corresponding hard band does not show such an increase.
Otherwise, within the errors the bins behave nearly identically as a
function of energy. 

\section{Data Analysis}
\subsection{General Spectral Fitting}

This section provides overall details on the spectral fitting.  Before
commencing any spectral fits, we corrected the data for the low-energy
contamination buildup \citep{plucinsky2003} and constructed a response
matrix appropriate for the specific detector and the off-axis position
of XY Ari. Spectra were extracted using the same apertures as defined
for the light curves (\S~\ref{obssect}).  All spectra were fit with
Sherpa v2.2.1 \citep{Freeman2001} and Xspec v11.2 \citep{Arnaud1996}
using ${\chi}^2$ data variance.  In each case, the continuum model was
fit first with the expected regions of line emission masked.  The line
masks were then removed and fits to the line regions were carried out.

Previous fits to X-ray spectra of XY Ari used absorbed thermal
bremsstrahlung models with kT=16, 17, and 30 keV (ASCA, Ginga, and
RXTE observations respectively, \citealt{Hellier1997,
EzukaIshida1998}).  Fits to the continuum of the ACIS data using a
thermal bremsstrahlung model, a power law model, and a blackbody model
all produced $\chi^{2}$ values of $\sim$1.1.  As our primary goal is
the study of line variations, we require a good but simple fit to the
continuum to define the lines.  We adopted an absorbed thermal
bremsstrahlung model not only to provide that simple fit but also for
comparison with previous results. 

The Chandra data do not constrain the bremsstrahlung temperature
because of the relatively high values for the temperature typical of
IPs and because of the declining effective area of the Chandra mirrors
above $\sim$5 keV.  We adopted a temperature of 30 keV as determined
from RXTE data because it has a broader bandpass than other satellites
such as Ginga or EXOSAT that have observed XY Ari.  

There are two different approaches for fitting lines, each with its
own merits: (i) place gaussians at the positions of known lines and
integrate the flux falling under the line; (ii) fit gaussians to
detected lines.  The first approach has the advantage of extracting
the most information with the least degree-of-freedom cost since the
line center is fixed.  This approach has the drawback that the placed
lines may not be statistically independent.  The second approach is
robust, but risks missing weak lines that fall below a pre-determined
significance threshold because of the extra degree-of-freedom cost.

We adopted the second approach because of the inherent resolution of
the ACIS detector.  We added as many gaussian lines as permitted by
the statistical improvement of the fit and the statistical
independence of the added line.  Based upon X-ray observations using
the Chandra High Energy Grating which shows narrow emission lines, the
lines are expected to be unresolved in the ACIS data
\citep{hellier2004,mukai2003}.  Therefore, each added gaussian had its
line width fixed at zero to mimic an unresolved line.

We calculated upper limits to any absent lines by adding a zero-width
gaussian at the line's position, then raising the gaussian
normalization until ${\Delta}{\chi}^2$ changed by the required amount.
In general, we quote 95\% upper limits, hence the required change was
4.0 for 1 parameter of interest (the line strength); for comparison
with values in the literature, we also use 1 ${\sigma}$ errors and
explicitly state when we do so.

The original Chandra data were obtained in two separate observations.
To reach the most sensitive detections for any line, we combined the
observations to maximize the signal-to-noise which will henceforth be
referred to as the 'combined data'.  We subsequently fit each
observation separately to detect any line strength variations over the
five-week separation between the observations.  We also broke each
observation into phase bins as detailed below.

\subsection{Overall Spectral Features}

\begin{figure*}
\begin{center}
\caption{The fit to the combined data set.  The emission lines have
deliberately not been included in the fit so the emission region is
more readily apparent.  The region from 5.2 to 7.2 keV, including the
fit, is shown expanded in the inset.}
\label{fig5}
\scalebox{0.4}{\rotatebox{-90}{\includegraphics{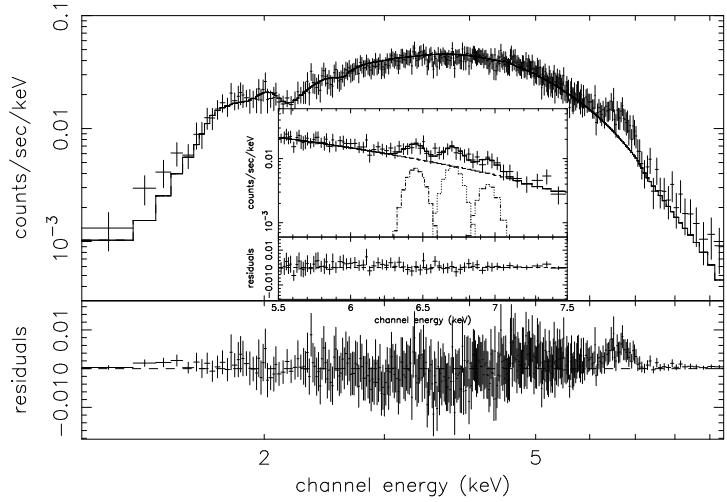}}}
\end{center}
\end{figure*}

From the combined data, we extracted the source and background spectra
in the same manner as described above for the light curves.  This
extraction was possible because XY Ari was in the same location on the
CCD to within a few pixels.  The total source spectrum was binned to
25 counts per channel and is shown in Figure~\ref{fig5}.  Lines near
6.4, $\sim$6.7, and 6.9 keV are visible; in addition, a weak line at
$\sim$8.2 keV is also marginally detected.  We henceforth use these
designations to label the lines.  We identify the strong lines as Fe I
K${\alpha}$ at 6.39 keV, the Fe XXV triplet complex in the 6.64-6.70
keV band, and Fe XXVI Ly${\alpha}$ 6.96 keV.

We then analyzed each data set separately.  For the separate
observations, we binned the spectra to 15 counts per channel.  We do
not show the spectra because they are visually similar to
Figure~\ref{fig5}.  Both observations separately show emission lines
at 6.4, 6.7, and 6.9 keV with significance $>$99\%.  The absorption
was also significantly higher during the second observation.  The
results from the fits to the combined and separate data sets are in
Table~\ref{tab1}.

\subsection{Phase-Resolved Fits and Results}

The count rate from XY Ari is sufficiently high to allow a study of
the time-resolved line emission, both at orbital and spin phases, but
sufficiently low to force considerations of the signal-to-noise per
phase bin.

Intermediate polars are generally not studied at orbital phases
because of the inherent blending of spin-dependent features.  In
\cite{Kamata93}, the authors published a plot of the orbital-dependent
behavior of the absorption column, reporting a sinusoidal-like
behavior.  Figure~\ref{fig8} shows the orbit-phased behavior of the
absorption column using 5 orbital phase bins for each observation.  We
confirm the variable behavior with orbital phase.

\begin{figure*}
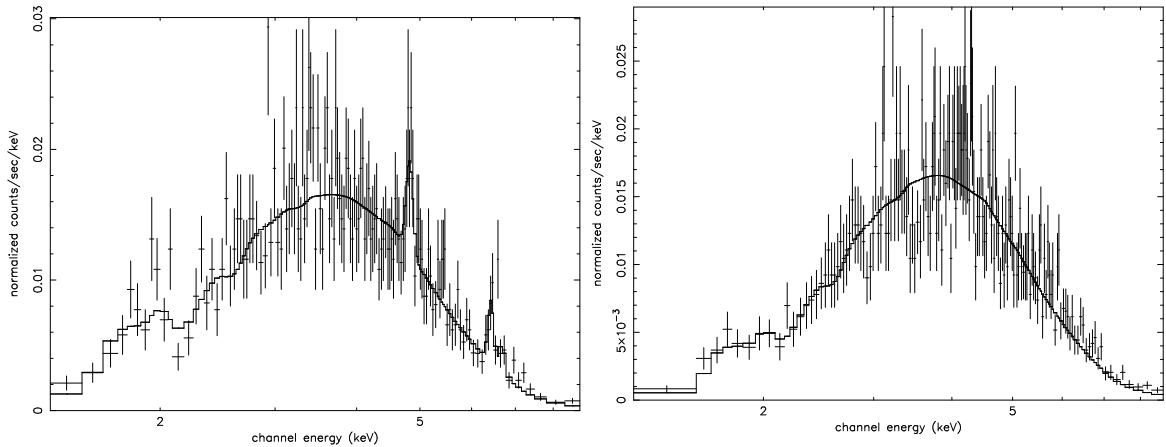

\begin{center}
\caption{\small Significantly different line emission in the 0.8-1.0
phase bin between the two observations.  (left) Observation 1: the 6.4
keV and 6.68 keV lines are clearly visible, as is the 4.8 keV line
(See \S{\ref{other}}). (right) Observation 2: no lines are visible.
The continuum temperature has been fixed at 30 keV as described in the
text.}
\label{fig7}
\scalebox{0.33}{\rotatebox{-90}{\includegraphics{Salinas.f5a.ps}}}
\scalebox{0.33}{\rotatebox{-90}{\includegraphics{Salinas.f5b.ps}}}
\end{center}
\end{figure*}

\begin{figure*}
\begin{center}
\caption{\small Changes in absorption during each orbit period.  Open
squares represent the first observation and filled blocks represent
the second observation. The data have been repeated twice for clarity.
This plot confirms an earlier study using {\it Ginga} \citep{Kamata93}.}
\label{fig8}
\scalebox{0.3}{\rotatebox{0}{\includegraphics{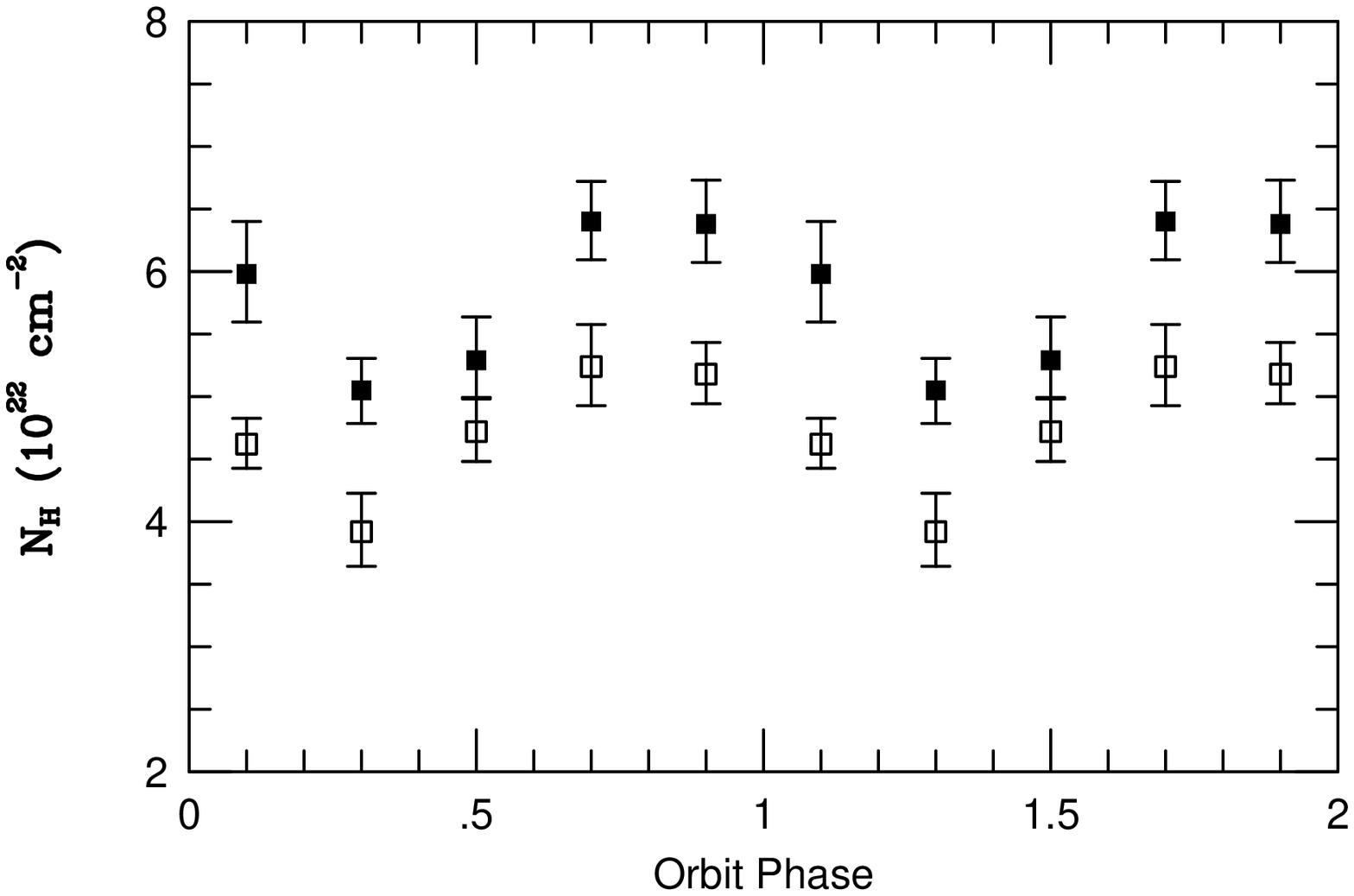}}}
\end{center}
\end{figure*}

For the spin-phased spectroscopy, we settled on two approaches.
First, based upon Figure~\ref{fig4}, spin phases $\sim$0.3-0.6 contain
a peak visible at all orbital phases.  We divided the data into 3 bins
phased on the spin period to study any orbital-dependent differences
directly.  Table~\ref{tab2} lists the results for the 3-bin fits.
Second, we divided the data into 5 bins again phased on the spin
period to push the phase resolution as much as possible;
table~\ref{tab3} lists the results for the 5-bin fits.  Examples of
two spectra from the 5-bin fits are shown in Figure~\ref{fig7} and
illustrate significant variations in line strengths.

\begin{figure*}
\begin{center}
\caption{\small Changes in absorption during each spin period.  Open
squares represent the first observation and filled blocks represent
the second observation. The data have been repeated twice for
clarity.}
\label{fig9}
\scalebox{0.3}{\rotatebox{0}{\includegraphics{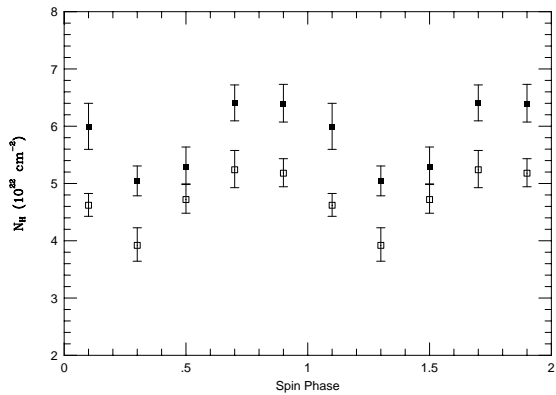}}}
\end{center}
\end{figure*}

Figure~\ref{fig9} shows the changes in the absorption column
throughout each observation and using the results from the 5-bin set
of data phased at the spin period.  The absorption is lowest at the
spin phases that correspond to low hardness ratios, and highest when
there is a peak in the hardness ratio.  This behavior matches the
model of additional matter along the line of sight at specific spin
phases, similar to a recent analysis of X-ray spectra of the IP PQ Gem
\cite{James02}; the added matter most likely originates from the
accretion stream or arc.  Alternatively, a second, soft emission
component could become visible during those phases and leading to a
smaller fitted value for the column density.  The spectral fit does
not require an additional emission component, however.

\begin{figure*}
\begin{center}
\caption{\small Changes in equivalent width during each spin period
for the 6.40, 6.68, and 6.96 keV lines.  Open symbols represent the
first observation and filled symbols the second.  Triangles represent
upper limits.  The data have been repeated twice for clarity.}
\label{fig10}
\scalebox{0.5}{\rotatebox{0}{\includegraphics{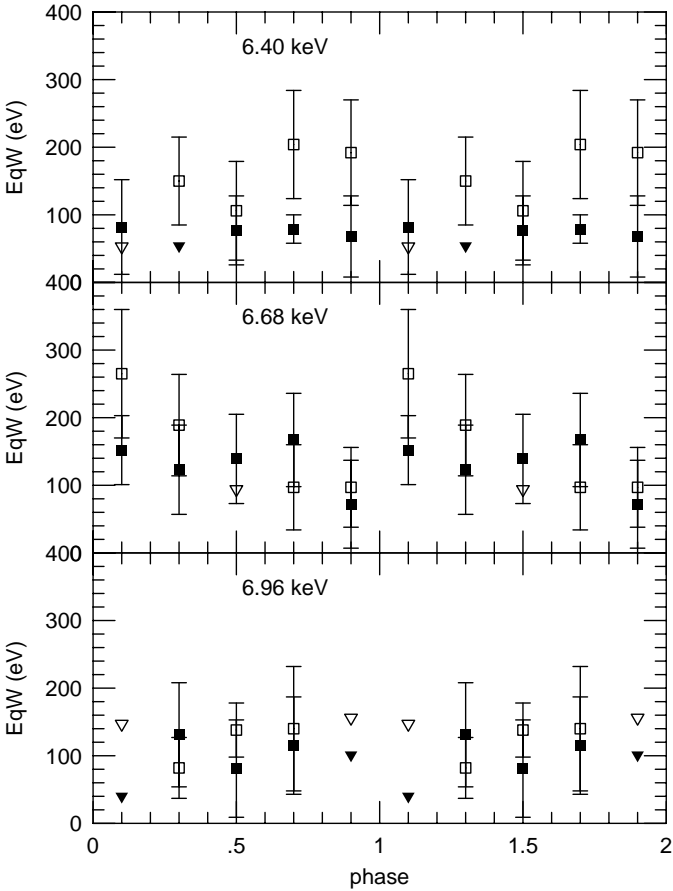}}}
\end{center}
\end{figure*}

The 6.4 keV line differs between the two observations and appears to
vary within an observation (Figure~\ref{fig10}).  A survival
statistics analysis confirms our impressions \citep{asurv}.  For the
6.4 keV line, the probability is low ($\sim$0.03) that the equivalent
widths from the two observations arise from the same parent
population.  The probability estimate is robust in that three
different survival tests (generalized Wilcoxon, Kaplan-Meier
estimator, log rank test) each yield similar probabilities (0.025,
0.033, 0.049, respectively).  The 3- and 5-bin mean Kaplan-Meier
estimates yield approximately a factor of 2 difference between the
observations over the 5-week gap for the 6.4 keV lines.  The 5-bin
mean values are 141$\pm$25 eV and 72$\pm$4 eV for the first and second
observations, respectively.  This represents a real detection of
differing line strengths in the 6.4 keV line.

Weak evidence exists for line variations in the 6.7 keV region
(Figure~\ref{fig10}).  The Kaplan-Meier estimates state that the mean
values are essentially identical at 148$\pm$31 and 131$\pm$15 eV for
the first and second observations, respectively; the 3-bin
Kaplan-Meier values are identical.  However, a fit of a constant or a
linear decline shows that a decline provides a marginally better fit
with ${\Delta}{\chi}^2$ decreasing by 2.7 which is significant at 90\%
for the extra degree-of-freedom.

For the 6.96 keV line, the line fluxes are constant with phase within
the errors; the Kaplan-Meier estimates of the means differ but are
just short of overlapping at 120$\pm$15 and 86$\pm$17 eV.  The
probability that the two samples were drawn from the same population
is $\sim$10-12\%.  Higher signal-to-noise spectra will be needed to
verify or refute any possible line variability.

\subsection{Unexpected Emission Lines} \label{other}

We also detected three emission lines in the first observation, each at
relatively large significance, that appear in the $\sim$3-6 keV band
and are not normally observed in astrophysical sources
(Table~\ref{tab4}).  The first line appears in the 0.2-0.4 phase bin
at 3.25 keV with $\sim$94\% significance, the second at 4.81 keV in
the phase 0.8-1.0 bin with $>$99\% significance (visible in
Figure~\ref{fig7}), and the third lies at 5.39 keV in the 0.0-0.2 phase
bin with a significance of $\sim$97\%.  We expended considerable
effort to understand their nature, as yet without success.

The Chandra calibration team has shown no line in the background at
any of the three detected line energies.  The XY Ari background data
do not show emission at these energies.  We applied current gain maps
to the files during their initial re-processing, so the lines are not
artifacts of the processing or the use of old gain values.  During the
orbit phases where the lines are seen, we filtered the data on energy,
isolating each line.  Images of the results show each line centered on
the position of XY Ari, which is counter to the expectation that the
lines arise from the background.

If we ignore the 0.5 keV difference in line energy for the moment, the
5.39 keV line could result from contamination from on-board
calibration sources, in which X-rays hitting the calibration devices
proceed to the detector and contaminate the data.  The Mn K$\alpha$
calibration source emits at 5.9 keV; the line is easily visible in
event histogram data (an instrument mode used for ACIS calibration),
but these data are collected {\it only} when ACIS is {\it not} in the
focal plane \citep{Calteam}.  The cal source might explain the line if
the line were present at all phases; the line, however, is {\it not}
so cooperative.  The line is essentially visible in a single phase bin.
	
Another possibility exists: the lines are red- or blue-shifted from
their rest energies.  Without a firm identification for each line, we
can not calculate velocities.  If we assume the lines are red-shifted
Fe K emission, the velocities involved are very large.  Adopting a
rest energy of 6.40 keV, we obtain velocities in the range of
0.2c-0.5c for the three lines, values well beyond the capabilities of
the potential well of a white dwarf.

Finally, charged particle events create emission lines in the
background spectrum from X-ray fluorescence of materials in the
detector.  If we take the fitted line centers as fluorescing lines,
then the 3.25 keV line might be attributed to Ar K${\alpha}$ (3.20
keV) and the 5.39 keV line to Cr Ka (5.41 keV).  The error on the line
center of the 3.25 keV line is sufficiently large that it encompasses
several ionization states of Ar.

The 4.81 keV line does not match any K line, as is expected for X-ray
fluorescence; possible matches regardless of mechanism are $^{53}$I
L${\gamma}$ (4.80 keV), $^{56}$Ba L${\beta}$ (4.83 keV), but both
lines are very weak \citep{xraydata}; the identification is not
plausible given the low abundance of these elements.  A more plausible
identification for the 4.81 keV line is Ca XIX with rest energy 4.823
keV.  The error on the line center of the 4.81 keV line easily
encompasses this rest energy and the abundance of Ca is sufficiently
high that the identification appears robust.  This is apparently the
first detection of Ca XIX in any CV.  

The 3.2 keV line, associated with Ar K${\alpha}$ above, also has
observational support: the {\it ASCA} spectrum of the intermediate
polar EX Hya shows emission from Ar XVII K${\alpha}$ and XVIII
K${\alpha}$ lines at 3.125 and 3.319 keV, respectively \citep{FI97}.  The
equivalent widths were measured as $\sim$20 eV each, broadly
consistent with unresolved emission in the ACIS spectrum if we assume
the two systems have similar radiation environments.

\section{Discussion}

XY Ari was in quiescence during the Chandra observation.  The
Einstein, Ginga, and quiescent RXTE luminosities L$_{\rm X}$ were
$\sim$1.7, $\sim$3.5, and $\sim$2.6, respectively, in units of
10$^{32}$ erg s$^{-1}$ in the 2-10 keV band
\citep{PattHalp90,Kamata93,HMB1997}.  These luminosities have been
corrected for distance, from the originally published values based on
a distance of $\sim$65 pc \citep{PattHalp90} to the more recent
estimate of 270$\pm$100 pc \citep{LDM}.  The mean Chandra luminosity
is 3.1${\pm}$0.1${\times}$10$^{32}$ erg s$^{-1}$.

We fit three different spectral data sets for XY Ari: the combined
data, the summed spectra for each observation separately, and the
phase-resolved spectra for each observation.  We discuss each in turn.

The lines detected in the combined data are at first glance similar to
the previous spectral studies using ASCA
\citep{Osborne,EzukaIshida1998}.  However, the Hellier et al. study
confidently detected only the 6.7 keV line; the 6.4 and 6.9 keV lines
in their table are really upper limits.  Similarly for Ezuka \&
Ishida: they list three lines in their table, but they detect the 6.4
keV line with 90\% confidence and report upper limits for the 6.7 and
7.0 keV lines.  In contrast, we detect three lines at $>$99\%
confidence in the combined data and in each observation separately
(Table~\ref{tab1}).  The differences in the detections among the
various studies may be attributed to variations in the signal-to-noise
of the corresponding spectra, but may also indicate variability in
line fluxes as a function of time.  If the lines vary, then
significant clues to the detailed accretion physics await detection of
the variations as a function of spin phase.

The presence of the 6.4, 6.7, and 6.9 keV lines provides constraints
on the ionization parameter $\xi$ = L$_{\rm X}$ / R$^2$ n, where R is
the size of the emitting region and n is the number density.  The 6.4
keV fluorescent line appears if ${\xi} <$2 \citep{KM82} and requires
an ionization stage of XVI or smaller.  The 6.7 and 6.9 keV lines,
however, arise from Fe XXV and XXVI, respectively, and require a
higher ionization state.  For the few phases where the lines are
equally detected, (e.g., observation 1, phases $\sim$0.3 and
$\sim$0.7), the 6.7/6.9 line strength ratio of $\sim$1 (range: 0.6 to
1.3) implies $\xi$ $\sim$3.5-4.  The upper limits on these lines,
particularly for observation 2, phase $\sim$0.1, require considerable
variation in the ionization state with spin phase.

Following \cite{BT99}, for temperatures below $\sim$12 keV, there is
essentially zero emission from Fe~XXVI Ly${\alpha}$.  The upper limits
on the line emission, particularly the more restrictive limits of
observation 2, require either a multi-temperature plasma or
significant blockage of the line emission; the latter inference
requires a contrived geometry to remove line emission without also
removing continuum emission.  Multi-temperature plasmas in the
accretion columns of IPs emerged from applications of the models of
\cite{Aizu73}; early spectral fits required only a single temperature.
Multi-temperature plasmas were first demonstrated from spectral fits
by \cite{Ishida94} using {\it ASCA} data.  The conclusion presented
here is therefore not new, but it confirms the physics using
phase-resolved spectroscopy of emission lines.  The emission from the
lines should be nearly co-located, in contrast to an alternative
explanation of the \cite{Ishida94} results, namely that the line
source regions were widely distributed.

The strength of the 6.4 keV line is proportional to the column density
of material \citep{kall95}: EqW $\sim$2.3 N$_{\rm 24}$ keV, where
N$_{\rm 24}$ is the column density in units of 10$^{24}$ cm$^{-2}$.
The measured mean values imply a factor of $\sim$2 change in the
column density.  The line strength is also related to the inclination
because of the increasing path length through the disk for
higher-inclination systems \citep{vanT}.  From figure 4 of the van
Teeseling et al. paper and for a specified inclination, a change in
the 6.4 keV equivalent width requires either a change in the
bremsstrahlung temperature or other processes, such as optical depth
effects or temperature inversions.

That the 6.4 keV fluorescence line is only detected at specific spin
phases will presumably tell us more about asymmetries in the cold iron
gas surrounding the accretion region.  Variations in ionization state
likely arise from aspect variations of the accretion region,
non-uniform filling factors of the absorbing gas, or optical depth
effects.  We know line emission occurs preferentially in directions
low optical depth; variations in optical depth could easily affect the
inferred phase-dependent effects reported here.  High wavelength- and
time-resolution observations will be necessary to sort out optical
depth effects.

We conclude with several speculative comments.  

The apparently transient behavior of the 4.8 keV line may be described as
flare-like.  Ca XIX is detected in solar flares (e.g., \citealt{Fludra99});
this line may represent the first such detection in CV.

The second observation showed marginal evidence for the presence of a
line at $\sim$6.25 keV.  If the `line' is actually a Compton shoulder
on the Fe K${\alpha}$ line arising from single-scattered photons, then
measurements of the column density of the scattering medium and its
metal abundance may be possible \citep{vanT,Watanabe2003}.

The spectral fit of the first observation is improved by a line at
$\sim$7.1 keV, perhaps the Fe I K${\beta}$ line; the
${\beta}:{\alpha}$ branching ratio is 17:150 \citep{Bam72}.
Similarly, the second observation detects a line at $\sim$8.2 keV,
possibly the Ly${\beta}$ line of Fe~XXVI.  The decreasing effective
area of the Chandra mirrors prevent robust detection of the continuum
and any possible line(s) above $\sim$7 keV, but provide two goals for
spectroscopy of IPs using future X-ray observatories since both lines
serve as temperature indicators.

\acknowledgements

We thank the referee for comments that materially improved the
presentation.  AS thanks everyone connected with the REU program for
support.  We thank Eli Beckerman for helping with aspects of the
software.  The research of EMS was supported by contract NAS8-39073 to
the Smithsonian Astrophysical Observatory.

\begin{table}
\begin{center}
\caption{Eclipse Times for XY Ari}
\label{ecltime}
\begin{tabular}{rll}
      & Relative &      \\
 Cycle & Time (days) & BJD \\ \hline
  0.0 & ~0.0 & 2451735.108505 \\
  1.0 & ~0.253472 & 2451735.361977 \\
151.0 & 38.157035 & 2451773.26554 \\
152.0 & 38.409875 & 2451773.51838 \\ \hline
\end{tabular}
\end{center}
\end{table}

\begin{table*}
\small
\begin{center}
\caption{Complete Data Sets: Combined/Weighted Data, First
Observation, Second Observation\tablenotemark{1}}
\label{tab1}
\begin{tabular}{ccccccc}
\hline\hline
         &Flux&Column &Line &Equivalent\\
	 & 2.0-10.0 keV  &Density& Energy & Width \\
Data Set &($10^{-12}$ ergs/$cm^{2}$/s)&($10^{22}cm^{-2}$) &(keV) &(eV)  \\
\hline
Combined &
$36.35^{+0.91}_{-0.88}$&$5.30^{+0.10}_{-0.10}$&$6.40^{+0.05}_{-0.03}$&108$\pm$38\\
	       	 &     &			     &$6.68^{+0.05}_{-0.08}$&106$\pm$46\\
		 &     &                             &$6.92^{+0.05}_{-0.05}$&117$\pm$51\\
		 &     &			     &$8.20^{+0.03}_{-0.18}$&90$\pm$51\\
\tableline
 First Observation&
$18.27^{+0.38}_{-0.41}$&$4.60^{+0.11}_{-0.11}$&$4.88^{+0.10}_{-0.11}$&48$^{+23}_{-40}$\\
                   &    &                     &$6.41^{+0.06}_{-0.02}$&133$^{+37}_{-46}$\\
		   &    &		      &$6.66^{+0.04}_{-0.06}$&106$^{+36}_{-44}$\\
		   &    &                     &$6.96^{+0.05}_{-0.08}$&103$^{+51}_{-64}$\\ \hline
 Second Observation&
$18.01^{+0.32}_{-0.30}$&$5.65^{+0.12}_{-0.12}$&$6.44^{+0.08}_{-0.06}$&76$^{+50}_{-36}$\\
		    &   &	              &$6.69^{+0.07}_{-0.04}$&157$^{+40}_{-50}$\\
		    &   &                     &$6.91^{+0.05}_{-0.08}$&104$^{+28}_{-54}$\\
		    &   &                     &$8.12^{+0.08}_{-0.10}$&104$^{+78}_{-71}$\\ \hline
\end{tabular}
\end{center}
\tablenotetext{1}{We adopted an absorbed thermal bremsstrahlung model
for the continuum with a fixed temperature of 30 keV.  All errors
are 95\% errors unless otherwise stated.}
\end{table*}

\begin{table*}
\small
\begin{center}
\caption{3-Bin Spin Phase Spectral Parameters\tablenotemark{1}}
\label{tab2}
\begin{tabular}{ccccccc}
\hline \hline
  & Unabsorbed  & Column & \multicolumn{4}{c}{Equiv. Width (eV): line at $\underline{~~~}$ keV}\\
Phase & Flux\tablenotemark{2} & Density\tablenotemark{3} & 6.40 & 6.68 & 6.96 & 7.20 \\ \hline
\multicolumn{7}{c}{Observation 1}\\
0.3-0.6 & 4.62$\pm$0.18 & 4.74$\pm$0.18 & 120$\pm$56 & ~98$\pm$51 & 147$\pm$75 & ~$<$98   \\
0.6-0.9 & 5.19$\pm$0.23 & 4.49$\pm$0.19 & 199$\pm$65 & ~97$\pm$51 & 125$\pm$73 & 119$\pm$93 \\
0.9-0.3 & 4.39$\pm$0.17 & 4.58$\pm$0.19 & 118$\pm$53 & 158$\pm$66 & ~53$\pm$51 & $<$88 \\
\multicolumn{7}{c}{Observation 2} \\
0.3-0.6 & 5.15$\pm$0.17 & 5.15$\pm$0.20 & ~47$\pm$42 & 116$\pm$47 & 121$\pm$63 & $<$86 \\
0.6-0.9 & 5.05$\pm$0.21 & 6.46$\pm$0.26 & ~91$\pm$43 & 172$\pm$57 & 162$\pm$59 & $<$73 \\
0.9-0.3 & 4.83$\pm$0.18 & 5.77$\pm$0.23 & 110$\pm$54 & ~68$\pm$54 & 120$\pm$72 & $<$87 \\ \hline
\end{tabular}
\tablenotetext{1}{We adopted an absorbed thermal bremsstrahlung model
for the continuum with a fixed temperature of 30 keV. Line widths have
been fixed at zero for all phases to approximate unresolved lines.
All errors are 95\%.}
\tablenotetext{2}{in units of 10$^{-12}$(ergs/cm$^{2}$/s) in the
energy range 2.0-10.0 keV.}
\tablenotetext{3}{in units of N$_{H}$ x 10$^{22}$ cm$^{-2}$.}
\end{center}
\end{table*}

\begin{table*}
\small
\caption{5-Bin Spin Phase Spectral Parameters\tablenotemark{1}}
\label{tab3}
\begin{tabular}{cccccc}
      & Unabsorbed & Column & \multicolumn{3}{c}{Equiv. Width (eV): line at $\underline{~~~}$ keV} \\
Phase & Flux\tablenotemark{2} & Density\tablenotemark{3} & 6.40 & 6.68
& 6.96  \\ \hline
\multicolumn{6}{c}{Observation 1} \\
0.0-0.2&2.98$\pm$0.09&4.62$\pm$0.23&~~~~~$<$53 &265$\pm$95&~~~~$<$147 \\
0.2-0.4&3.07$\pm$0.14&3.92$\pm$0.19&150$\pm$65 &189$\pm$75&~82$\pm$45 \\
0.4-0.6&3.26$\pm$0.14&4.72$\pm$0.25&106$\pm$73 &~~~~~$<$94&138$\pm$40 \\
0.6-0.8&3.08$\pm$0.09&5.24$\pm$0.30&204$\pm$80 &~97$\pm$63&140$\pm$92 \\
0.8-1.0&3.22$\pm$0.09&5.18$\pm$0.27&192$\pm$78 &~97$\pm$59&~~~~$<$156 \\
\multicolumn{6}{c}{Observation 2} \\
0.0-0.2&3.42$\pm$0.10&5.98$\pm$0.39&~82$\pm$61 &152$\pm$84 &~~~~~$<$40 \\
0.2-0.4&3.23$\pm$0.11&5.05$\pm$0.24&~~~~~$<$54 &123$\pm$66 &131$\pm$77 \\
0.4-0.6&3.63$\pm$0.13&5.29$\pm$0.33&~77$\pm$51 &139$\pm$66 &~81$\pm$72 \\
0.6-0.8&3.38$\pm$0.08&6.40$\pm$0.31&~79$\pm$53 &167$\pm$69 &115$\pm$72 \\
0.8-1.0&3.47$\pm$0.10&6.38$\pm$0.32&~68$\pm$60 &~72$\pm$65 &~~~~$<$101 \\ \hline
\end{tabular}

\tablenotetext{1}{We adopted an absorbed thermal bremsstrahlung model
for the continuum with a fixed temperature of 30 keV. Line widths were
fixed at zero for all phases.  Errors on the equivalent widths are 1
${\sigma}$ for comparison with values from the literature; other
errors are 95\%.}

\tablenotetext{2}{in units of 10$^{-12}$(ergs/cm$^{2}$/s) in the
energy range 2.0-10.0 keV.}
\tablenotetext{3}{in units of N$_{H}$ x 10$^{22}$ cm$^{-2}$.}
\end{table*}

\begin{table*}
\caption{Unexpected Emission Lines in Observation 1\tablenotemark{1}}
\label{tab4}
\begin{tabular}{llcrrcl}
 Line  & Phase & EquivW &                    &         & Upper Limits at & Line \\
Center & Observed   & (eV)   & ${\Delta}{\chi}^2$ & Signif. & Other phases\tablenotemark{2} & Identity \\ \hline
3.25$\pm$0.04 & 0.2-0.4 & 44 & ~5.5 & 94\% & /36/$\cdots$/27/24/77/ &
Ar K${\alpha}$ 3.20 keV? \\
4.81$\pm$0.03 & 0.8-1.0 & 99 & 15.3 & $>$99\% & /41/81/37/30/$\cdots$/
& Ca XIX 4.832 keV? \\
5.39$\pm$0.04 & 0.0-0.2 & 85 & ~7.8 & 97\% & /$\cdots$/85/86/40/117/ & no id? \\ \hline
\end{tabular}
\tablenotetext{1}{~Lines were fit using a gaussian of zero width to
approximate an unresolved emission line.}
\tablenotetext{2}{~Upper limits at other phases: the 95\% limits are
  defined at phases /0.0-0.2/0.2-0.4/0.4-0.6/0.6-0.8/0.8-1.0/ and in
  that order.}
\end{table*}

\end{document}